\documentstyle[sprocl]{article}

\def\be{\begin{equation}}
\def\ee{\end{equation}}
\def\bea{\begin{eqnarray}}
\def\eea{\end{eqnarray}}

\begin{document}

\title{NON-ABELIAN STOKES THEOREM AND QUARK CONFINEMENT IN QCD\footnote{Invited talk given at International Symposium on Quantum Chromodynamics (QCD) and Color Confinement (Confinement 2000), Osaka, Japan, 7-10 Mar 2000,
$\&$ at 30th International Conference on High-Energy Physics (ICHEP 2000), Osaka, Japan, 27 Jul - 2 Aug 2000 (ICHEP2000) (to be published in the Proceedings (World Scientific, Singapore)).}
}

\author{K.-I. KONDO}

\address{Department of Physics, Faculty of Science, \\ Chiba University, Chiba 263-8522, Japan\\E-mail: kondo@cuphd.nd.chiba-u.ac.jp} 




\maketitle\abstracts{ 
To understand the Abelian dominance and magnetic monopole dominance in low-energy 
QCD, we rewrite the non-Abelian Wilson loop into the form which is written in terms of 
its Abelian components or the 't Hooft-Polyakov tensor describing the magnetic 
monopole.  This is peformed by making use of a version of non-Abelian Stokes theorem.  
We propose a modified version of the maximal Abelian (MA) gauge.  By adopting the 
modified MA gauge in QCD, we show that the off-diagonal gluons and Faddeev-Popov 
ghosts acquire their masses through the ghost--anti-ghost condensation due to four 
ghost interaction coming from the gauge-fixing term of the modified MA gauge.  The 
asymptotic freedom of the original non-Abelian gauge theory is preserved in this 
derivation.
}

\section{Introduction}
In this talk we discuss quark confinement in QCD based on the Wilson criterion, i.e., the 
area (decay) law of the Wilson loop.  The main subject of this talk is to understand the 
Abelian dominance in low-enegy QCD which was found by Suzuki and 
Yotsuyanagi\cite{SY90} based on Monte Carlo simulations of lattice gauge theory 
under the maximal Abelian (MA) gauge proposed by Kronfeld et al.\cite{KLSW87}.  
The Abelian dominance was predicted by Ezawa and Iwazaki\cite{EI82} immediately 
after the proposal of the Abelian projection by 't Hooft\cite{tHooft81}.  The Abelian 
dominance and the subsequent magnetic monopole dominance is quite important to 
understand quark confinement from the viewpoint of the dual superconductor picture\cite{Nambu74} of the QCD vauum, since the condensation of magnetic monopole can 
lead to the dual superconductivity based on the electro-magnetic duality argument.  
\par
In this talk we give a pedagotical introduction to a version of non-Abelian Stokes 
theorem (NAST).  This version of the NAST clearly shows (in the operator level) the 
relationship between the Wilson loop and its Abelian components or the magnetic 
monopole. 
This fact is very suggestive of the possible Abelian and monopole dominance after 
taking the expectation value of the Wilson loop.  
\par
For the Abelian dominance to hold in low-energy region of QCD, it is sufficient to show 
that the off-diagonal gluons become massive and they can be in a sense neglected in the 
low-energy region (although the latter statement is not necessarily true as shown in 
this talk).
In order to really understand the Abelian dominance, we need to know the mechanism 
of the mass generation for the off-diagonal gluons.  
We propose a modified version of the MA gauge.  Then we show that mass generation 
can be understood as a consequence of taking the (modified) MA gauge. We will give two 
kinds of explanations.  One is given by a rather formal argument based on a novel 
reformulation of the gauge theory (as a perturbative deformation of a topological 
quantum field theory (TQFT)) proposed by the author\cite{KondoII}.  Another is to 
examine the explicit form of the gauge fixing term in the MA gauge\cite{KS00a}.  We 
give a prediction on the off-diagonal gluon mass.

\section{A version of non-Abelian Stokes theorem (NAST)}

We make use of a version of non-Abelian Stokes theorem (NAST) which is obtained by 
making use of the idea of Dyakonov and Petrov\cite{DP}.

{\bf Theorem:}\cite{KT99}  {\it For a closed loop $C$, we define the non-Abelian 
Wilson loop operator by
\begin{equation}
  W_C[{\cal A}] = {\rm tr} \left\{ P \exp \left[ ig \oint_C dx^\mu {\cal A}_\mu(x) 
\right] \right\}/{\rm tr}(1) ,
\end{equation}
where $P$ is the path-ordered product.
Then it is rewritten as
\begin{eqnarray}
  W_C[{\cal A}] &=& \int d\mu_C(\xi) \exp \left[ ig \oint_C dx^\mu 
a_\mu^{\xi}(x) \right] 
\\
&=& \int d\mu_C(\xi) \exp \left[ ig \int_S d\sigma^{\mu\nu} f_{\mu\nu}^{\xi}(x) 
\right] ,
\end{eqnarray}
where
\begin{equation}
 a_\mu^{\xi}(x) := \langle \Lambda | {\cal A}_\mu^\xi(x) |\Lambda \rangle , 
\quad
 {\cal A}_\mu^\xi(x) = \xi^\dagger(x) {\cal A}_\mu(x) \xi(x) + {i \over g} 
\xi^\dagger(x) \partial_\mu \xi(x) ,
\label{eq:acomp}
\end{equation}
and
\begin{equation}
  f_{\mu\nu}^{\xi} := \partial_\mu a_\nu^{\xi} - \partial_\nu a_\mu^{\xi} .
\end{equation}
Here $|\Lambda \rangle$ is the highest-weight state of the representation defining 
the Wilson loop and the measure $d\mu_C(\xi)$ is the product measure 
$d\mu_C(\xi)=\prod_{x \in C}d\mu(\xi(x))$ on $G/\tilde H$ with the maximal 
stability group $\tilde H$. The maximal stability group is the subgroup leaving the 
highest-weight state invariant (up to a phase factor) and depends on the $G$ and the 
representation in question.}
\par
For $G=SU(2)$, the $\tilde H$ is given by the maximal torus subgroup 
$H=U(1)$ irrespective of the representation. Hence $G/\tilde H=CP^1=F_1$.  For 
$G=SU(N) (N \ge 3)$, however, $\tilde H$ does not necessarily agree with 
$H=U(1)^{N-1}$ depending on the representation.  For $G=SU(3)$, all the 
representations can be classified by the Dynkin index $[m,n]$.  If $m=0$ or $n=0$,   
$\tilde H=U(2)$ and $G/\tilde H=CP^2$.  On the other hand, when $m\not=0$ and 
$n\not=0$, $\tilde H=U(1)^2$ and $G/\tilde H=F_2$.  Here $CP^n$ is the complex 
projective space and $F_n$ the flag space.  This NAST is obtained by making use of the 
generalized coherent state.  For details, see Perelomov or Feng, Gilmore and Zhang.
\par
 For the fundamental representation, the expression (\ref{eq:acomp}) is greatly 
simplified as
\begin{equation}
 \langle \Lambda | (\cdots) |\Lambda \rangle = 2 {\rm tr}[{\cal H}(\cdots)] ,
\quad
 {\cal H}={1 \over 2}{\rm diag}\left({N-1 \over N},{-1 \over N}, \ldots, {-1 \over N} 
\right) .
\end{equation}
Therefore, $a_\mu= {\cal A}_\mu^3$ for $G=SU(2)$, and $a_\mu= {\cal 
A}_\mu^3+{1 \over \sqrt{3}} {\cal A}_\mu^8$ for $G=SU(3)$.
This implies that the non-Abelian Wilson loop can be expressed by the diagonal 
(Abelian) components.  This is suggestive of the Abelian dominance in the expectation 
value of the Wilson loop.
\par
The monopole dominance in the Wilson loop is also expected to hold as shown follows.
We can rewrite $f_{\mu\nu}^{\xi}$ in the NAST as  
\begin{equation}
 f_{\mu\nu}^{\xi} = \partial_\mu[n^A {\cal A}_\nu^A] - \partial_\nu[n^A {\cal 
A}_\mu^A] - {1 \over g} f^{ABC} n^A \partial_\mu n^B \partial_\nu n^C ,
\end{equation}
where 
\begin{equation}
 n^A(x) T^A = \xi(x) {\cal H} \xi^\dagger(x) .
\end{equation}
The  $f_{\mu\nu}^{\xi}$ is invariant under the full $G$ gauge transformation as well 
as the residual $H$ gauge transformation.  (Indeed, we can write a manifestlly gauge 
invariant form, see ref.\cite{HU99}.)
This is nothing but the 't Hooft-Polyakov tensor of the non-Abelian magnetic monopole, 
if we identify $n^A(x)$ with the unit vector of the elementary Higgs scalar field in the 
gauge-Higgs theory:
\begin{equation}
 n^A(x) \leftrightarrow \hat \phi^A(x) := \phi^A(x)/|\phi(x)| .
\end{equation}
This implies that $n^A(x)$ is identified with the composite scalar field and plays the 
same role as the scalar field in the gauge-Higgs model, even though QCD has no 
elementary scalar field.  This fact could explain why the QCD vacuum can be dual 
superconductor due to magnetic monopole condensation.
By introducing the magnetic monopole current $k$ by 
$k := \delta {}^*f$, we have another expression,
\begin{eqnarray}
  W_C[{\cal A}] 
= \int d\mu_C(\xi) \exp \left[ ig (N, k^{\xi}) \right] ,
\quad
N := \Delta^{-1}{}^* dS ,
\end{eqnarray}
where $\Delta$ is the Laplacian and $S$ is the area two-form on the surface spanned 
by the Wilson loop $C$.
Hence, the Wilson loop can also be expressed by the magnetic monopole current 
$k_\mu$.
\par
In the case of $SU(2)$, the Wilson loop in an arbitrary representation (characterized by 
$J=1/2,1,3/2,\cdots$) is written in the form,
\begin{eqnarray}
  W_C[{\cal A}] &=& \int d\mu_C(\xi) \exp \left[ ig J \oint_C dx^\mu 
a_\mu^{\xi}(x) \right] ,
\end{eqnarray}
where
\begin{equation}
 a_\mu^{\xi}(x) := {\rm tr}\{ \sigma_3 [\xi^\dagger(x) {\cal A}_\mu(x) \xi(x) + i 
g^{-1}\xi^\dagger(x) \partial_\mu \xi(x)] \} .
\end{equation}

\par

\section{The modified MA gauge}
When we calculate the expectation value 
$\langle W_C[{\cal A}] \rangle_{YM}$
of the Wilson loop, we must specify the procedure of the gauge fixing.
In order to incorporate the magnetic monopole in the non-Abelian gauge theory without 
the elementary scalar (Higgs) field, we adopt the modified MA gauge to define the 
gauge-fixed QCD.  The gauge fixing (GF) and the Faddeev-Popov (FP) term of the 
modified MA gauge is given by
\begin{equation}
  S_{GF+FP} = \int d^4x \ i \delta_B \bar \delta_B \left[ {1 \over 2}A_\mu^a(x) 
A^{\mu}{}^a(x) - {\alpha \over 2} i C^a(x) \bar C^a(x) \right] ,
\label{MAg}
\end{equation}
where $\delta_B$ ($\bar \delta_B$) is the BRST (anti-BRST) transformation.
The special case $\alpha=-2$ was discussed by several papers\cite{KondoII,KondoIV,KondoVI}.
The modified MA gauge fixing term which is the BRST and anti-BRST exact and FP 
conjugation\cite{antiBRST} invariant has a hidden 
$OSp(4|2)$ supersymmetry\cite{KondoII}.  
Due to this supersymmetry, the dimensional reduction of Parisi-Sourlas\cite{PS79} 
type takes place.  
\par
\par
For simplicity, we discuss only the SU(2) case.  For SU(3) case, see ref.\cite{KS00a}.
\begin{eqnarray}
  S_{GF+FP}' &=&   \int d^4x  \Biggr\{ 
B^a D_\mu[a]^{ab}A^\mu{}^b+ {\alpha \over 2} B^a B^a
\nonumber\\
&&+ i \bar C^a D_\mu[a]^{ac} D^\mu[a]^{cb} C^b
- i g^2 \epsilon^{ad} \epsilon^{cb} \bar C^a C^b A^\mu{}^c A_\mu^d 
\nonumber\\
&&
+ i \bar C^a g  \epsilon^{ab} (D_\mu[a]^{bc}A_\mu^c) C^3 
\nonumber\\
&&- \alpha  g \epsilon^{ab} i B^a \bar C^b C^3 
+ {\alpha \over 4} g^2 \epsilon^{ab} \epsilon^{cd} \bar C^a \bar C^b C^c C^d \Biggr\} .
\label{GF4}
\end{eqnarray}
Integrating out the $B^a$ field leads to
\begin{eqnarray}
  S_{GF+FP}' &=&   \int d^4x  \Biggr\{ 
-{1 \over 2\alpha}(D_\mu[a]^{ab}A^\mu{}^b)^2  
\nonumber\\
&&+i \bar C^a D_\mu[a]^{ac} D^\mu[a]^{cb} C^b
- i g^2 \epsilon^{ad} \epsilon^{cb} \bar C^a C^b A^\mu{}^c A_\mu^d 
\nonumber\\
&&+ {\alpha \over 4} g^2 \epsilon^{ab} \epsilon^{cd} \bar C^a \bar C^b C^c C^d 
\Biggr\} .
\label{GF5}
\end{eqnarray}
A crucial difference of the modified MA gauge from the conventional Lorentz 
gauge is the necessity of four ghost interactions for renormalizability.  Even for 
$\alpha=0$, the four ghost interaction term is induced through radiative corrections 
due to the existence of the $c\bar c AA$ vertex.

\section{Deformation of a TQFT and the dimensional reduction}
\par
The author\cite{KondoII} has proposed a novel reformulation of the gauge theory, i.e., a 
perturbative deformation of a topological quantum field theory (TQFT).  Here a part 
extracted from the GF+FP term (\ref{MAg}) is identified with a TQFT.  By making use 
of the NAST within this reformulation, we have calculated the expectation value of the 
Wilson loop and obtained the area law\cite{KondoII,KondoIV}.  In this calculation, we 
have utilized the dimensional reductin from TQFT$_4$ to NLSM$_2$.  Actual 
calculations are performed by 1) the instanton calculus (in the dilute gas 
approximation)\cite{KondoII} and by 2) the large N expansion (in the leading 
order)\cite{KT99}.  The static interquark potential is obtained as
\begin{equation}
  V(R) = \sigma R - {N^2-1 \over 2N}{\alpha_s \over R}f(R) + const.,
\end{equation}
where $f(R) \rightarrow 1$ as $R \rightarrow 0$.  The second term of the potential 
comes from the perturbative deformation part where the coefficient $(N^2-1)/(2N)$ for 
SU(N) was obtained by Prosperi\cite{Prosperi99}.  The string tension $\sigma$ is 
obtained for the fundamenal representation as
\begin{equation}
  \sigma = m^2 \exp \left[ -|\alpha| {2\pi^2 \over g^2} \right] ,
\end{equation}
In the instanton calculus, the mass dimension $m$ is required by the dimensional 
reasons for defining the measure of the instanton size.  On the other hand, $m$ is 
equal to the mass of NLSM$_2$ and calculable in the large N expansion.
The coupling constant $g$ should run through the renormalization of the potential 
$V(R)$.  However, this calculation is not so easy.  We look for other route.

\section{Ghost self-interaction and dynamical mass generation}
\par
Integrating out off-diagonal field components $(A_\mu^a, B^a, C^a, \bar C^a)$ in 
Yang-Mills theory in the MA gauge, we can obtain an effective Abelian gauge theory 
written in terms of only the diagonal components $(a_\mu^i, B^i, C^i, \bar C^i, 
B_{\mu\nu}^i)$.  This theory called the Abelian projected effective gauge theory 
(APEGT) is regarded as a low-energy effective theory (LEET) of QCD.  The coupling 
constant of APEGT has the $\mu$(renormalization-scale) dependence governed by the 
$\beta$ function which is the same as the original Yang-Mills 
theory.\cite{KondoI,KS00b} This reflects the asymptotic freedom of the original non-Abelian gauge theory.  The other RG functions and the anomalous dimensions have 
been calculated recently\cite{KS00b}.
\par
In the MA gauge, the renormalizability requires the existence of four ghost interactions.  
The modified MA gauge determines the strength of four ghost interaction where the 
modified MA gauge is obtained from the viewpoint of pursuing the maximal symmetry, 
namely, BRST, anti-BRST, FP conjugation and $OSp(4|2)$ supersymmetry.
The attractive four ghost interaction leads to two types of ghost--anti-ghost 
condensations,\cite{KS00a,Schaden99} 
\begin{equation}
  \epsilon^{ab} \langle i \bar C^a(x) C^b(x) \rangle \not=0 ,
\quad
  \epsilon^{ab} \langle i \bar C^a(x) C^b(x) \rangle = {v \over 16\pi} \not=0 .
\end{equation}
In the condensed vacuum, the ghost-gluon 4-body interaction, 
\begin{equation}
    -i g^2 \epsilon^{ad} \epsilon^{cb} \bar C^a C^b A^\mu{}^c A_\mu^d ,
\end{equation}
leads to a mass term of the off-diagonal gluons, 
\begin{equation}
    -i g^2 \epsilon^{ad} \epsilon^{cb} 
\langle \bar C^a C^b \rangle A^\mu{}^c A_\mu^d  
=   {1 \over 2} g^2 \langle i \bar C^c C^c \rangle A^\mu{}^a A_\mu^a ,
\end{equation}
Thus this condensation leads to the mass for the off-diagonal gluons 
\begin{equation}
  m_{A}^2 = g^2 \langle i  \bar C^a(x) C^a(x) \rangle > 0 .
\end{equation}
On the other hand, the off-diagonal ghost (and anti-ghost) acquires the 
mass,\cite{KS00a}
\begin{equation}
  m_C^2 = \alpha g^2 \langle i \bar C^a(x) C^a(x) \rangle ,
\end{equation}
through the four ghost interaction,
\begin{eqnarray}
 && {\alpha  \over 4} g^2 \epsilon^{ab} \epsilon^{cd} \bar C^a \bar C^b C^c C^d 
=   {\alpha  \over 2} g^2 (i \epsilon^{ab} \bar C^a C^b)^2
=  {\alpha  \over 2} g^2 (i \bar C^a C^a)^2
\nonumber\\
&&\rightarrow    \alpha  g^2 \langle i \bar C^a C^a \rangle
i \bar C^b C^b .
\end{eqnarray}
Note that the introduction of the explicite mass term spoils the renormalizability.  It 
can be shown that the diagonal gluons remain massless.
 The mass obtained in this way provides the scale which is comparable to the QCD scale 
$\Lambda_{QCD}$.\cite{KS00a}  This result is consistent with the lattice simulations performed
by Amemiya and Suganuma\cite{AS99}.
The dynamical mass generation for the off-diagonal components strongly supports the 
Abelian dominance in low-energy (or long-distance) QCD.  
\par
At least for $G=SU(2)$, the Lagrangian in the modified MA gauge has a novel 
(continuous) global symmetry, $SL(2,R)$ as found by Schaden\cite{Schaden99}.  Then the mass generation 
can be considered as a spontaneous breaking of this symmetry from $SL(2,R)$ to the 
non-compact Abelian subgroup corresponding to the ghost number charge $Q_c$.  This 
mechanism of mass generation can be called the dynamical Higgs mechanism, since 
QCD has no elementary scalar field.  The associated massless Nambu-Goldstone 
particles can not be observed, since they have zero norms due to the extended quartet 
mechanism.\cite{KO79} In these analyses, we have assumed that the vacuum satisfies 
the physical condition,
\begin{equation}
Q_B|0 \rangle = 0, \quad Q_c |0 \rangle = 0, \quad
\bar Q_B|0 \rangle = 0   .
\end{equation}
\par
For $G=SU(3)$, the ghost condensation scenario for mass generation of the off-diagonal 
components can be applied and leads to two different masses for off-diagonal gluons; 
two of them are heavier than the remaining four off-diagonal gluons, e.g., 
\begin{equation}
m_{A^1}=m_{A^2}=\sqrt{2}m_{A^4}=\sqrt{2}m_{A^5}=\sqrt{2}m_{A^6}=\sqrt{2}m_{A
^7}.\end{equation}

\section{Discussion}

\par
Finally, we raise the problems to be solved in the future investigations.
\par
1.  All the results obtained above are invariant under the residual $U(1)^{N-
1}$ Abelian gauge symmetry.  However, they may depend on the gauge parameter 
$\alpha$ of the MA gauge.  In the recent work, $\alpha$ was determined by requiring 
the $\mu$ independence of the effective potential of the order parameter of the ghost 
condensation as
\begin{equation}
 \alpha=b_0/N=11/3 .
\end{equation}
\par
2. The proof of renormalizability of QCD in the (modified) MA gauge has not yet been 
given when the ghost condensation takes place.  In the absence of ghost condensation, 
the proof of renormalizability was given 15 years ago cite{MLP85}.  
\par
3. For $SU(3)$, no one has shown the existence of a global symmetry whose 
spontaneous breaking leads to the mass generation of off-diagonal fields (through the 
ghost condensation).  Hence the relationship between the mass generation and the 
spontaneous symmetry breaking is not yet understood in a satisfactory level.
\par
4.  It is important to show how the dynamical mass of the off-diagonal gluons is related 
to the mass of the dual gauge field in the dual Abelian gauge theory (e.g., dual 
Ginzburg-Landau theory\cite{dGL}) which is expected to be another LEET of QCD.

\section*{Acknowledgments}
The author would like to thank the organizers, Hideo Suganuma and Hiroshi Toki for 
inviting me to give a talk in this workshop.
This work is supported in part by
the Grant-in-Aid for Scientific Research from the Ministry of
Education, Science and Culture (10640249).


\end{document}